\documentstyle{article}

\newtheorem{Definition}{Definition}
\newtheorem{Theorem}{Theorem}
\newtheorem{Lemma}{Lemma}

\newcommand{\proof}{\noindent {\bf Proof: }}
\newcommand{\qed}{$\Box$}

\begin{document}
\pagestyle{plain} 

\title{\Large{\bf{Quantum lower bounds for the set equality problems}}}

\linespread{0.8}
\author{\small{Gatis Midrij\= anis}
\\
\\\small{University of Latvia,}
\\\small{29 Raina boulevard, Riga, Latvia}
\\\small{gatis@zzdats.lv}}

\date{}
\maketitle

\linespread{1}
\begin{abstract}
    \small{
    The set equality problem is to decide whether two sets $A$ and $B$ are equal or
    disjoint, under the promise that one of these is the case.
    Some other problems, like the Graph Isomorphism problem, is
    solvable by reduction to the set equality problem. It was an open problem to find any $w(1)$
    query lower bound when sets $A$ and $B$ are given by quantum
    oracles with functions $a$ and $b$.

    We will prove $\Omega(\frac{n^{1/3}}{\log^{1/3} n})$
    lower bound for the set equality problem when the set of the preimages
    are very small for every element in $A$ and $B$.
    }
\end{abstract}

\thispagestyle{empty}

\linespread{1} \topmargin 0.8cm \textwidth 12cm \textheight 19.5cm
\marginparwidth 4.5cm

\section{\large{\textbf{Introduction, Motivation and Results}}}
\label{sec:Intro}

 The Shor's integer factoring
quantum algorithm provides exponential speed-up over the best
known classical algorithm. This motivates to search other quantum
algorithms with great speed-up. However, proving quantum lower
bounds for such problems is not trivial, for example, proving the
exponential quantum lower bound for NP-Complete problems will
imply $P \neq NP$.

One of the problems quantum computer could have an exponential
speed-up over classical computer is the Graph Isomorphism problem.
One way to attack this problem could be by the reduction to the
set equality problem. Notice the sets of all permutations over
vertexes for given graphs. If these sets are equal, then there is
an isomorphism between the graphs, but if there is not isomorphism
between graphs, then these sets are strictly disjoint.

Denote the set $\{1, 2, ..., n\}$ by $[n]$.
\begin{Definition}
Let $a:[n]\mapsto [m]$ and $b:[n]\mapsto [m]$ be a functions. Let
$A$ be a set of all $a's$ images $A = \{a(1), a(2), ..., a(n)\}$
and $B = \{b(1), b(2), ..., b(n)\}$. There is a promise that
either $A = B$ or $A \cap B = \O$.

Call \textbf{the general set equality} problem to distinguish
these two cases.
\end{Definition}

Finding quantum query lower bound for general set equality problem
was posed an open problem by Shi\cite{ShiColl}.

We will show that Ambainis' \cite{Ambainis} adversary method imply
$\Omega(\sqrt{n})$ lower bound for the general set equality
problem. The proof uses the possibility to have many preimages for
some image. However, graph theorists think that the Graph
Isomorphism problem, when graphs are promised not to be equal with
themselves by any nonidentical permutation, still is very complex
task. Now reduction lead us to the set equality where $a$ and $b$
are one-to-one functions.

\begin{Definition}
Call the general set equality problem to be a \textbf{one-to-one
set equality} problem if $a(i) \neq a(j)$ and $b(i) \neq b(j)$ for
all $i \neq j$.
\end{Definition}

The proof that worked for the general set equality problem does
not work for one-to-one set equality problem, because it uses that
fact that there can be very many preimages for any element of the
sets. However, we will prove lower bound for a problem between
these problems.

\begin{Definition}
Call the general set equality problem to be a \textbf{f(n) set
equality} problem if $|a^{-1}(x)| = O(f(n))$ and $|b^{-1}(x)| =
O(f(n))$ for all images $x \in a([n]) \cup b([n])$ and for some
function $f$.
\end{Definition}

We will prove $\Omega(\frac{n^{1/3}}{\log^{1/3}n})$ lower bound
for the $log(n)$ set equality problem.

The first result for lower bounds of the set equality like problem
was done by Aaronson \cite{Aaronson}. He showed $\Omega(n^{1/6})$
lower bound for so called set comparison problem: to decide
whether two sets are equal or disjoint on a constant fraction of
elements. He also assumed that both $a$ and $b$ are one-to-one
functions. In this paper, we will study lower bound of problem
when these sets $A$ and $B$ are strictly disjoint or equal,
however $a$ and $b$ is not a one-to-one.

\section{Preliminaries}
\label{sec:Preliminaries}

\subsection{\normalsize{\textbf{Quantum Query algorithms}}}
\label{sec:Preliminaries:upper}

 The most popular model of quantum computing is a
query (oracle) model where the input is given by a black box. For
more details, see a survey by Ambainis~\cite{AmbSurv} or textbook
by Gruska~\cite{Gruska}. In this paper we are able to skip them
because our proof will be built on reduction to solved problems.

One of the most amazing quantum algorithms is a Grover's search
algorithm. It shows how a given $x_1 \in \{0, 1\}, x_2 \in \{0,
1\}, ..., x_n \in \{0, 1\}$ to find the $i$ that $x_i = 1$ with
$O(\sqrt{n})$ queries.

This algorithm can be generalized to so called amplitude
amplification \cite{AmplAmpl}. Using amplitude amplification one
can make good quantum algorithms for many problems till the
quadratic speed-up over classical algorithms.

By straightforward use of amplitude amplification we get a quantum
algorithm with $O(\sqrt{n})$ queries for the general set equality
problem and a quantum algorithm with $O(n^{1/3})$ queries for the
one-to-one set equality problem.

\subsection{\normalsize{\textbf{Quantum query lower bounds}}}
\label{sec:Preliminaries:lower}

 There are two main approaches to get good quantum
lower bounds. The first is Ambainis' \cite{Ambainis} quantum
adversary method. The other is lower bound by polynomials
introduced by Beals et al. \cite{Beals} and substantially
generalized by Aaronson~\cite{Aaronson} and Shi~\cite{ShiColl}.
Although explicitly we will use only Ambainis' method, main result
we will get by a reduction to problem, solved by polynomials'
method.

The basic idea of adversary method is that, if we can construct
relation $R \subseteq A \times B$, where $A$ and $B$ consisting of
0-instances and 1-instances and there is a lot of ways how to get
from an instance in $A$ to an instance in $B$ that is in the
relation and back by flipping various variables, then query
complexity must be high.

\begin{Theorem}
\label{AThm} \cite{Ambainis} Let $f(x_1, ..., x_N)$, be a function
of n \{0, 1\}-valued variables and $X, Y$ be two sets of inputs
such that $f(x) \neq f(y)$ if $x \in X$ and $y \in Y$. Let $R
\subset X \times Y$ be such that
\begin{itemize}
    \item For every $x \in X$, there exist at least $m$ different $y
        \in Y$ such that $(x, y) \in R$.
    \item For every $y \in Y$, there exist at least $m'$ different
    $x \in X$ such that $(x, y) \in R$.
    \item For every $x \in X$ and $i \in \{1, ..., n\}$, there are at most $l$ different
    $y \in Y$ such that $(x, y) \in R$ and $x_i \neq y_i$.
    \item For every $y \in Y$ and $i \in \{1, ..., n\}$, there are at most $l'$ different
    $x \in X$ such that $(x, y) \in R$ and $x_i \neq y_i$.
\end{itemize}
Then, any quantum algorithm computing $f$ uses
$\Omega(\sqrt{\frac{m m'}{l  l'}})$ queries.
\end{Theorem}

\subsection{\normalsize{\textbf{The collision problem}}}
\label{sec:Preliminaries:col}

Finding $w(1)$ quantum lower bound for the collision problem was
an open problem since 1997. In 2001 Scott Aaronson \cite{Aaronson}
solved it showing polynomial lower bound. Later his result was
improved by Yaoyun Shi \cite{ShiColl}. Newly Shi's result was
extended by Samuel Kutin \cite{Kutin} and by Andris Ambainis
\cite{AmbCol} in another directions.

Below is exact formulation of collision problem due to
Shi\cite{ShiColl}.
\begin{Definition}
Let $n > 0$ and $r \geq 2$ be integers with $r|n$, and let a
function of domain size n be given as an oracle with the promise
that it is either one-to-one or r-to-one. Call the
\textbf{r-to-one} collision problem the problem to distinguishing
these two cases.
\end{Definition}

Shi~\cite{ShiColl} showed quantum lower bound for r-to-one
collision problem.
\begin{Theorem} ~\cite{ShiColl} \label{shi}
    Any error-bounded quantum algorithm to solve r-to-one collision must
    evaluate the function $\Omega((n/r)^{1/3})$ times.
\end{Theorem}

\section{\large{\textbf{Results}}}
\label{sec:Main}

\subsection{\normalsize{\textbf{Lower bound for the general set equality
problem}}} \label{sec:Main:general}

\begin{Theorem}
    Any quantum algorithm which solves the general set equality
    problem makes $\Omega(\sqrt{n})$ queries.
\end{Theorem}
\emph{Proof.} Simple use of Ambainis' Theorem~\ref{AThm}. Since
Ambainis' Theorem~\ref{AThm} deals with boolean functions, we will
modify any quantum algorithm that solves the general set equality
problem to an algorithm, that computes boolean function.

We will prove this theorem even in a restricted case, when
functions returns only two values, let say $0$ and $1$. So we have
a problem, given two functions $a:n\mapsto \{0, 1\}$ and
$b:n\mapsto \{0, 1\}$ answer either the sets $A = \{a(1), ...,
a(n)\}$ and $B = \{b(1), ..., b(n)\}$ are equal or disjoin under
the promise that one of this is the case.

Let $f:\{0, 1\}^{2n}\mapsto \{0, 1\}$ be partially defined
function, such that
    \begin{center}
    \begin{math}
    f(a_1, a_2, ..., a_n, b_1, b_2, ..., b_n) =
    \left\{
    \begin{array}{ll}
        1, & \hbox{if $\{a_1, ...,
a_n\} = \{b_1, ..., b_n\}$;} \\
        0, & \hbox{if $\{a_1, ...,
a_n\} \cap \{b_1, ..., b_n\} = \O$.} \\
    \end{array}
    \right.
    \end{math}
    \end{center}
It is easy to see, that if we can solve a general set equality
problem, we can compute this function with constant slowdown, too.

Let construct the relation $R$ from Ambainis' Theorem~\ref{AThm}
with $X = \{0^n1^n\}$ and $Y= \{0^i10^{n-i-1}1^i01^{n-i-1} : 0
\leq i < n\}$ as follows: $$R = X \times Y = \{(0^n1^n,\
0^i10^{n-i-1}1^i01^{n-i-1}) : 0 \leq i < n\}.$$

One can check that $R$ is well defined and $m = n$, $m' = 1$, $l =
1$ and $l' = 1$. Thus any quantum algorithm computing $f$ makes
$\Omega(\sqrt{n})$ queries. \qed

\subsection{\normalsize{\textbf{Lower bound for the log(n) set equality
problem}}} \label{sec:Main:start}

Now we will prove the main result in this paper.
\begin{Theorem} \label{main}
    Any error-bounded quantum algorithm which solves the log(n) set equality
    problem makes $\Omega(\frac{n^{1/3}}{\log^{1/3}n})$ queries.
\end{Theorem}
\proof

To prove Theorem~\ref{main} we will reduce r-to-one collision
problem to the log(n) set equality problem. We are given function
$f:[n]\mapsto [m]$, under promise to be either r-to-one or
one-to-one and $r = \lceil\log n\rceil$ and $r|n$. We randomly
choose two sets $A$ and $B$ such that $|A| = |B| = n/2$ and $A
\cup B = [n]$ and $A \cap B = \O$. Denote $A' = f(A)$ and $B' =
f(B)$. It is obviously that if $f$ is one-to-one then $A' \cap B'
= \O$.

If $f$ is r-to-one then the situation is more complicate. In the
next subsection we will prove that with big probability holds that
$A'$ and $B's$ includes all images of f, thus $A' = B' = f([n])$.

Let the functions $a$ and $b$ from Theorem~\ref{main} be the same
as $f$ but domain for $a$ is $A$ and domain for $b$ is $B$.

Denote the set of all preimages of x in the set $A$ by
${f^{-1}}_A(x) = f^{-1}(x) \cap A$. Since $|f^{-1}(x)| = r$ for
every $x \in f([n])$, it is clear that also $|a^{-1}(x)| = O(\log
n)$ and $|b^{-1}(x)| = O(\log n)$ for every $x \in f([n])$.

So with constant probability we get the log(n) set equality
problem with domain size $n/2$. Now Theorem~\ref{shi} implies
Theorem~\ref{main}. \qed

\subsection{\normalsize{\textbf{Reduction}}}
\label{sec:Main:red}

\begin{Lemma}
\label{lemma:red}

From all possible divisions of a set $[n]$ into two equal sized
parts $A$ and $B$ such that $A \cap B = \O$, only few of them are
such that for some $x \in f([n])$ there is no preimage either in
$A$ or $B$.
\end{Lemma}

\proof

Total count of all (possibly uniform) divisions are
$${\mathcal{C}}^{n/2}_{n} = \frac{n!}{(n/2)!(n/2)!}.$$

Total count of such divisions is at most the count of images
($n/r$) multiplied by count of divisions where one fixed $x \in
f([n])$ has no preimage either in $A$ or $B$.

Assume that all preimages of $x$ is in $A$, thus $B$ has not any
of them. Number of ways how we can choose residual elements is
$${\mathcal{C}}^{n/2-r}_{n-r} = \frac{(n-r)!}{(n/2-r)!(n/2)!}.$$
Analysis of an opposite assumption is similar, so probability to
choose division which is bad on $x$ is
$$\frac{2{\mathcal{C}}^{n/2-r}_{n-r}}{{\mathcal{C}}^{n/2}_{n}}=\frac{2(n-r)!(n/2)!(n/2)!}{(n/2-r)!(n/2)!n!} = \frac{2(n/2)(n/2-1)(n/2-2)...(n/2-r+1)}{n(n-1)(n-2)...(n-r+1)}=$$
$$=\frac{n/2-1}{n-1}\frac{n/2-2}{n-2}...\frac{n/2-r+1}{n-r+1} \leq (\frac{1}{2})^{r-1}.$$

So probability to choose bad division for any $x \in f([n])$ is at
most $$(\frac{1}{2})^{r-1}\frac{n}{r}=\frac{2n}{2^rr}.$$ Since $r
=\lceil \log n \rceil$ $$\frac{2n}{2^rr} \leq \frac{2}{\lceil \log
n \rceil}$$ which is small for large $n$.

 \qed

\section{\large{\textbf{Conclusion}}}
\label{conclusion}

Finding lower bound for the set equality problem is one of the
most challenging today's task in theory of quantum query lower
bounds. We have solved this problem partially. One can argue that
to solve the set equality problem can be easier when functions are
promised to be with a small range of preimages for all images. Our
paper shows that the difference between the general set equality
problem and the $\log n$ set equality problem is very small,
respectively $\Omega(\sqrt{n})$ and
$\Omega(\frac{n^{1/3}}{\log^{1/3} n})$. This enforce opinion that
quantum computer probably cannot solve the one-to-one set equality
problem with only polylogarithmic number of questions.

\section{\large{\textbf{Acknowledgments}}}
I am very grateful to Andris Ambainis for introducing me with
quantum algorithms and also with this problem as well as comments
during writing this paper.

Research supported by Grant No.01.0354 from the Latvian Council of
Science, and  Contract IST-1999-11234 (QAIP) from the European
Commission.

\end{document}